\documentclass[12pt,twoside]{article}
\usepackage{fleqn,espcrc1}

\usepackage{epsfig}
\usepackage[figuresright]{rotating}

\newcommand{\AmS}{{\protect\the\textfont2
  A\kern-.1667em\lower.5ex\hbox{M}\kern-.125emS}}

\title{Time Scales in Spectator Fragmentation
\thanks{Work supported by the European Community under contract ERBFMGECT-950083}
}

\author{C.~Schwarz
	\thanks{Work supported by the Deutsche Forschungsgemeinschaft 
	under contract Schw510/2-1}
	\address{Gesellschaft  f\"ur  Schwerionenforschung, D-64291 Darmstadt, Germany\\
	$^b$ Istituto di Scienze Fisiche, Universit\`{a} degli Studi
     	di Milano and I.N.F.N., I-20133 Milano, Italy\\
	$^c$ Department of Physics and
     	Astronomy and National Superconducting Cyclotron Laboratory,
     	Michigan State University, East Lansing, MI 48824, USA\\
	$^d$ Dipartimento di Fisica dell' Universit\`{a} and I.N.F.N.,
     	I-95129 Catania, Italy\\
	$^e$ Forschungszentrum Rossendorf, D-01314 Dresden, Germany\\ 
	$^f$ Institut f\"ur Kernphysik,
     	Universit\"at Frankfurt, D-60486 Frankfurt, Germany\\
	$^g$ Soltan Institute for Nuclear Studies, 00-681 Warsaw, 
     	Hoza 69, Poland
	},
S.~Fritz$^a$,
R.~Bassini$^b$,
M.~Begemann-Blaich$^a$
	\thanks{Work supported by the Deutsche Forschungsgemeinschaft 
	under contract Be1634/1-1},
S.J.~Gaff-Ejakov$^c$,
D.~Gourio$^a$,
C.~Gro\ss$^a$,
G.~Imm\'{e}$^d$,
I.~Iori$^b$,
U.~Kleinevo\ss$^a$
	\thanks{Present address: Fachbereich Physik, Bergische Universit\"at,
	D-42119 Wuppertal, Germany},
G.J.~Kunde$^c$
	\thanks{Present address: Dept. of Physics, Yale University, 
	New Haven CT 06512, USA},
W.D.~Kunze$^a$,
U.~Lynen$^a$,
V.~Maddalena$^d$,                   
M.~Mahi$^a$,
T.~M\"ohlenkamp$^e$,
A.~Moroni$^b$,
W.F.J.~M\"uller$^a$,
C.~Nociforo$^d$,                    
B.~Ocker$^f$,
T.~Odeh$^a$,
F.~Petruzzelli$^b$,
J.~Pochodzalla$^a$
	\thanks{Work supported by the Deutsche Forschungsgemeinschaft 
	under contract Po256/2-1}
	\thanks{Present address: Max-Planck-Institut f\"ur Kernphysik,
D-69117 Heidelberg, Germany},
G.~Raciti$^d$,
G.~Riccobene$^d$,                   
F.P.~Romano$^d$,                   
A.~Saija$^d$,                       
M.~Schnittker$^a$,
A.~Sch\"uttauf$^f$
	\thanks{Present address: Max-Planck-Institut f\"ur Physik,
	D-80805 M\"unchen, Germany},
W.~Seidel$^e$,
V.~Serfling$^a$,
C.~Sfienti$^d$,                     
W.~Trautmann$^a$,
A.~Trzcinski$^g$,
G.~Verde$^d$,
A.~W\"orner$^a$,
Hongfei~Xi$^a$,
and B.~Zwieglinski$^g$
}  

\begin{document}

\maketitle

\begin{abstract}
Proton-proton correlations and correlations of p-$\alpha$,
d-$\alpha$, and t-$\alpha$ from spectator decays following
$ ^{197}$Au+$^{197}$Au collisions at 1000 AMeV have been measured with
an highly efficient detector hodoscope. The constructed correlation
functions indicate a moderate expansion and low breakup densities
similar to assumptions made in statistical multifragmentation models.
In agreement with a volume breakup rather short time scales were deduced
employing directional cuts in proton-proton correlations.
\end{abstract}


\section{INTRODUCTION}

Understanding the simultaneous formation of many fragments helps 
to gain an insight into multifragmentation. Densities lower than ground state
density of nuclei are a prerequisite of statistical models describing 
multifragmentation \cite{Bondorf951,Gross972}. 
\begin{figure}[tbh]
\begin{minipage}[h]{6.5cm}
\includegraphics*[width=6cm]{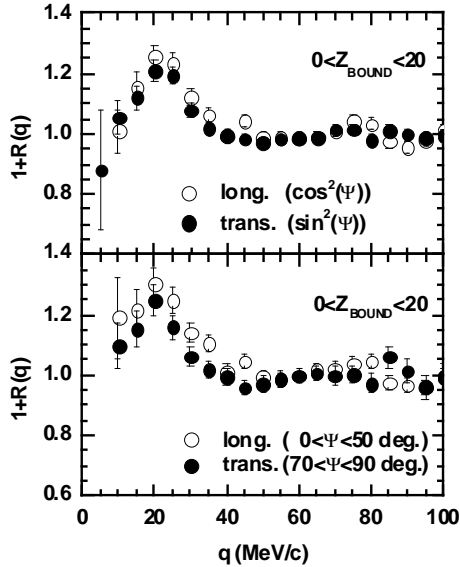}
\end{minipage}
\begin{minipage}[b]{10cm}
\parbox{9.5cm}
{
\vspace{-5cm}
\caption{\small \label{ppcorr}
Proton-Proton correlation-functions for central collisions. 
The top panel shows directional selections generated with 
harmonic weights, the bottom panel shows directional selection of pairs of protons,
whose angular ranges of the angle between relative and sum momentum are indicated 
in the figure.}
}
\end{minipage}
\end{figure}
While static statistical models 
form fragments in an expanded volume breakup, the dynamic statistical model
\cite{Friedman837,Friedman901} combines an expansion with surface emission 
until the volume breakup of the remaining source occurs. 
Interferometry-type methods are widely considered
as a valuable tool in determining the space-time extent of such a source and recently,
spectator remnants of the reaction Au + Au at 1 AGeV incident energy were found
to break up at densities, considerable lower than ground state density \cite{Fritz995}.
In that analysis a instantaneous volume breakup was assumed.
In such a breakup the mean time difference between the
emission of two fragments is assumed to be short, due to fluctuations of the breakup time. 

In this article we report on results of correlation measurements of protons from
the target spectator following
the collisions of Au + Au at 1 AGeV incident energy. Coincidences between protons
were used to construct correlation functions. 
The results are found to be consistent
with low breakup densities with values close to those assumed in the statistical
multifragmentation models and short emission time differences with values
close to those anticipated for volume breakup.

\section{EXPERIMENT}

Targets, consisting of $25$ mg/cm$^{2}$ of $^{197}$Au were irradiated
by an 1 AGeV Au beam delivered by the Schwerionen-Synchrotron (SIS) at GSI in
Darmstadt. 
For the results, presented here, we employed one multi-detector
hodoscope, consisting of a total of 96 Si-CsI(Tl) telescopes in closely packed
geometry. The hodoscope covered an angular range $ \Theta _{lab} $ from 122$ ^{0} $
to 156$ ^{0} $ with the aim of selectively detecting the products of the
target-spectator decay. Each telescope consisted of a 300 \mbox{$\mu$m}  Si detector
with \mbox{30 x 30 mm$^{2}$} active area, followed by a 6 cm long CsI(Tl) scintillator
with photodiode readout. The distance to the target was 60 cm. The products
of the projectile decay were measured with the time-of-flight wall of the ALADIN
spectrometer \cite{Schuettauf961} and the quantity $ Z_{BOUND} $ was determined
event-by-event. $ Z_{BOUND} $ is defined as the sum of the atomic numbers
$ Z_{i} $ of all fragments with $ Z $$ _{i} $$ \geq 2 $. It reflects
the variation of the charge of the primary spectator system and serves as a
measure for the impact parameter.

\section{DATA ANALYSIS}

\subsection{Correlation functions}

The correlation functions were constructed dividing the spectrum of relative
momenta of two coincident particles by the spectrum of pairs from
different events. One observes at a relative
momentum of $ q\approx 20 $ MeV/c a maximum of the correlation function,
which is dominated by the attractive singlet scattering between two protons. 
The height of the maximum is inversely related to the relative
distance between two protons and hence, the diameter of the source 
for simultaneous emission \cite{Koonin779,Pratt871}.
\begin{figure}[b]
\begin{minipage}[h]{9.5cm}
\includegraphics*[angle=-90,scale=0.35]{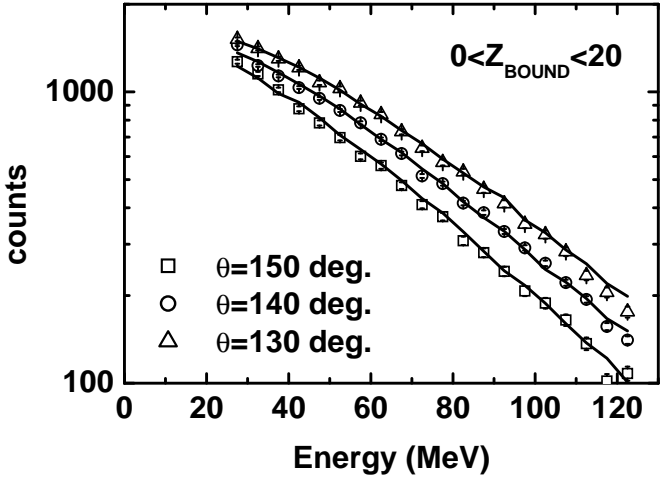}
\end{minipage}
\begin{minipage}[b]{7cm}
\parbox{6.5cm}
{
\vspace{-6cm}
\caption{\small \label{spectra1}
Energy spectra of protons at angles of 130, 140, and 150 deg. in the 
laboratory system.}
}
\end{minipage}
\end{figure}
Different emission times
of the protons cause an elongation of the source in the direction of the hodoscope
\cite{Lisa933}.
\begin{table}[b]
\newcommand{\m}{\hphantom{$-$}}
\newcommand{\cc}[1]{\multicolumn{1}{c}{#1}}
\renewcommand{\tabcolsep}{1.6pc} 
\begin{tabular}{cccccc}
\hline
{$Z_{BOUND}$}&  $T$  & $P_{bounce}$& $P_{Coulomb}$ & v$_{stop}$ &Mult.\\
             & (MeV) &   (MeV/c)   &   (MeV/c)   &     (c)     &       \\
\hline
0-20 & 26 & 105 &  80 &  0.020 & 6.1 \\
20-40 & 23 &  85 & 120 & 0.015 & 6.2 \\
40-60 & 19 &  80 & 140 & 0.010 & 5.8 \\
60-79 & 16 &  80 & 160 & 0.005 & 4.2 \\
\hline
\end{tabular}\\[2pt]
\caption{\small \label{source}
Source parameters for the MC-calculation of the protons: 
given are the temperature $T$, the Gaussian 
momentum width of the transversal movement of the source, the maximum 
momentum due to the Coulomb repulsion, the Gaussian width of the proton-multiplicity 
spectrum, and the velocity of the source in beam direction v$_{stop}$.}
\end{table}
Since two protons in the perpendicular direction are closer,
they suffer due to the Pauli blocking a stronger suppression. We selected pairs
of protons with directional cuts on the angle $ \Psi  $ between the sum momentum
\mbox{$\bf P$} and the relative momentum \mbox{$\bf q$}. The choice of the sum momentum as 
reference direction is motivated by the tilted source elongation with respect
to the beam axis. The notion for longitudinal (${\bf P} \parallel {\bf q}$) and transversal
(${\bf P} \perp {\bf q}$) correlation functions, therefore, differs from what is 
commonly used in high energy physics. Instead of using cuts on the angle $\Psi$
between sum momentum and relative momentum we employed for the first time
a harmonic weight for each event. We used $\cos^2(\Psi)$ and $\sin^2(\Psi)$ as 
weight for the longitudinal and transversal correlation function, respectively. 

\begin{figure}[tbh]
\begin{centering}
\includegraphics*[angle=-90,scale=0.85]{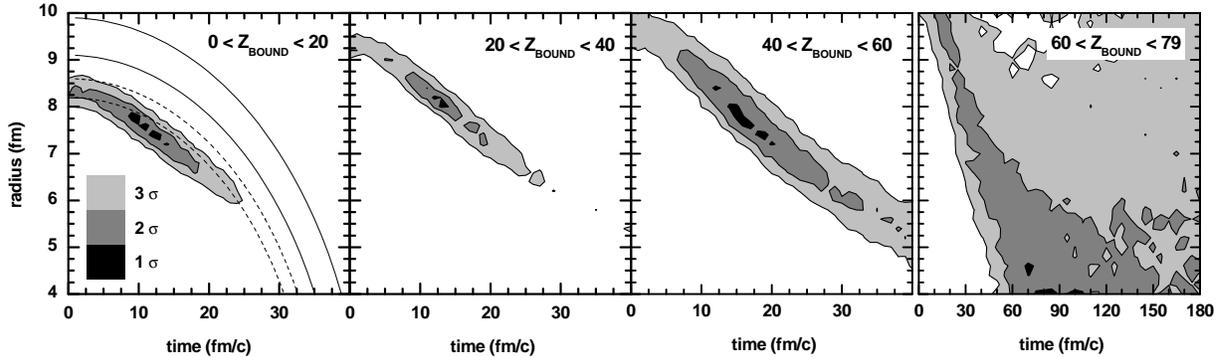}
\caption{\small \label{maps}
$\chi^2$ distributions as a function of radius and emission time of the 
emitting source for four impact parameter ranges indicated by the $Z_{BOUND}$ ranges
in the panels. The lines in the left panel are explained in the text.}
\end{centering}
\end{figure}
The proton-proton correlation-functions are shown in Fig. \ref{ppcorr} for the most 
central impact parameter bin. We used protons with energies $E>20$ Mev in order
to prevent distortions from sequential decay \cite{Schwarz991}. 
The correlation functions show a pronounced peak due to the final state interaction
at \mbox{$q=20$ Mev/c} relative momentum. The longitudinal selection exhibits at lower
relative momenta, \mbox{$q<50$ MeV/c}, correlations which are $\approx 5\%$ stronger 
compared with the transversal selection. For a source with Radius \mbox{$R=7$ fm} one can 
roughly estimate from the uncertainty relation the relative momenta 
where antisymmetrization effects should take place as 
\mbox{$q=\sqrt{{\bf q}^2}$} \mbox{$<\sqrt{3}*\hbar/r \approx 50$ MeV/c}.  
The bottom panel of Fig. \ref{ppcorr} serves as comparison between the two methods for
generating directional cuts. Here, pairs of protons were selected according to the
orientation between sum momentum and relative momentum as it was done in analyses 
performed before \cite{Lisa933}. 
Employing harmonic weights gives the advantage to use the full data set for 
one directional selection, while for the hard cuts less than half of the events are used.
In detector setups with limited angular acceptance a hard transversal cut limits
the statistic for large relative momenta and makes the normalization more difficult.
This eg. can be seen from the larger error bars of the transversal correlation function in 
the bottom panel of Fig. \ref{ppcorr} for relative momenta \mbox{$q>60$ MeV/c}. Here, 
many of the proton pairs with large transversal relative momenta miss the hodoscope.       

\subsection{Monte-Carlo simulations}
The analysis of the p-p correlation functions
was performed with the Koonin-Pratt formalism \cite{Koonin779,Pratt871}.
Particles were chosen to be emitted randomly from the volume of an uniform sphere and 
their velocities were sampled according to a Maxwell-Boltzmann 
distribution. An additional velocity component was added, which simulates the 
Coulomb repulsion dependent on the location of the particle within the source.
The strength of this component was chosen recursively, in order to agree with the later 
extracted source radii.
From the measured proton-energy spectra (Fig. \ref{spectra1}) one can observe that 
the emission probability of the source is non isotropic. We modeled this anisotropy 
by assuming a sideward movement (bounce) of the source perpendicular to the beam, 
which causes a reaction plane. It influences slightly 
the correlation functions by the event mixing technique which we used: the choice of 
a pair of protons from events with different reaction planes causes an additional small 
relative momentum between the momenta. Therefore, the relative momentum spectrum of 
the mixed events is slightly shifted to larger values compared to the spectrum of
the coincident proton pairs. The correlation function then shows a slightly decreasing 
slope. 
Since our experiment lacks an azimuthal symmetric solid angle coverage, we were not able
to determine the reaction plane and to rotate the different events in the same plane 
to avoid these effects\cite{Kaempfer931}.
We rather preferred to introduce the reaction plane in the Monte-Carlo calculations.
In order to simulate the hit probability of the hodoscope for different bounce values, 
we adjusted the multiplicity of the event such that it resembles the measured 
proton-multiplicity spectrum after applying the experimental filter. The extracted source 
parameters are given in Tab. \ref{source}.

\section{RESULTS}

We varied the radius of an uniform density distribution and the 
Gaussian emission time of the protons. 
The time distribution is guided by the idea of a statistically independent emission 
of different emitters within the source, rather than the decay of an excited state 
(see eg. \cite{Boal861}). The simulated correlation functions were used to perform 
a $\chi^2$-test algorithm in the relative momentum region of 
\mbox{$10 \le q \le 35$ MeV/c}. The results are presented in Fig. \ref{maps}. 
The grey scale of the shading denotes within which standard deviation the space-time 
point is. The minima in Fig. \ref{maps} show approximately constant source radii and 
surprisingly short emission times. For the most peripheral bin a radius and emission 
time could due to the low statistics of the correlation function not be deduced. 
The experimental correlation functions (symbols) and the simulations (lines) for the 
best $\chi^2$ are shown 
\begin{figure}[bth]
\begin{minipage}[h]{11cm}
\includegraphics*[scale=0.38,angle=-90]{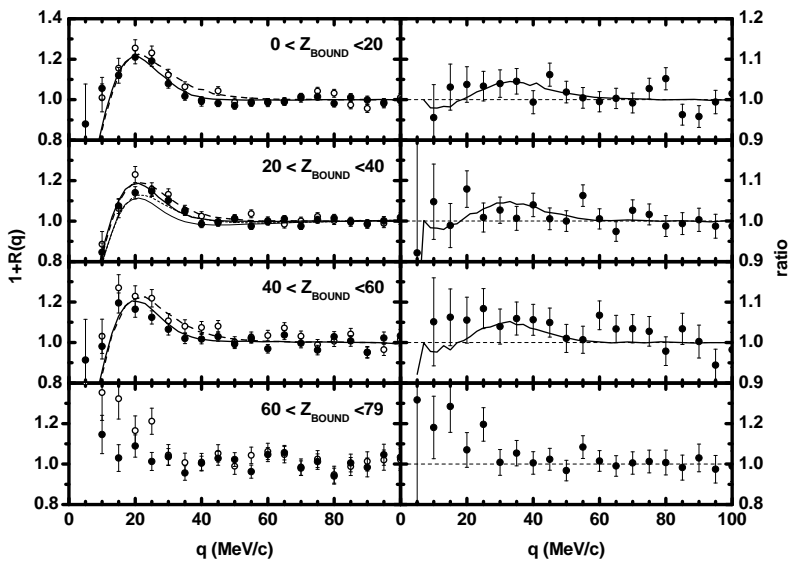}
\end{minipage}
\begin{minipage}[b]{4.5cm}
\parbox{5cm}
{
\vspace{-3cm}
\caption{\small \label{datasim}
The left panel shows experimental longitudinal (open symbols) and 
transversal (closed symbols) correlation functions and results of MC-simulations (lines). 
The right panel compares the their experimental ratios (symbols) with the results of the 
MC-simulation (lines).}
}
\end{minipage}
\end{figure}
in Fig. \ref{datasim}, left panel.  
They agree well with their simulations. The minimum \mbox{$\chi^2$} values per degree of 
freedom
are within the range of \mbox{$1.3 \le \chi^2/d.o.f. \le 2.4$}. 
The right panel in Fig. \ref{datasim} presents the ratios between longitudinal and 
transversal correlation functions. One recognizes the weak enhancement of about 5\% 
above unity (dashed line) of data (symbols) and simulations (solid line) 
for relative momenta $q<50$ MeV/c. For low values, \mbox{$q<20$ MeV/c},  
this enhancement becomes suppressed by the Coulomb repulsion between the protons.
Although the sensitivity of the directional cuts extends up to relative momenta of
$q<50$ MeV/c, we restricted our fit region to lower values of \mbox{$10 \le q \le 35$ MeV/c}. 
In this region the shape of the simulated correlation functions describes well the data.
The ratios in the right panel of Fig. \ref{datasim} 
agree, however, similar well for the larger relative momenta.
Using a fit region of $10 \le q \le 50$ MeV/c yields radii within 
\mbox{$\Delta R=0.2$ fm} and emission times within \mbox{$\Delta \tau=2$ fm/c} 
to those results extracted from the narrower relative momentum region.  
\begin{figure}[b]
\begin{minipage}[h]{8.5cm}
\includegraphics*[scale=0.3,angle=-90]{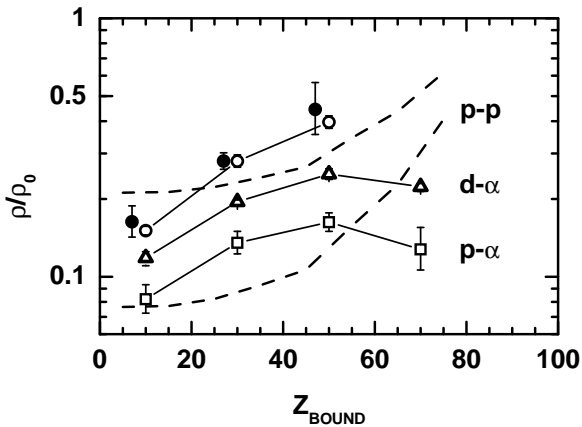}
\end{minipage}
\begin{minipage}[b]{8cm}
\parbox{7.5cm}
{
\vspace{-2cm}
\caption{\small \label{densities}
Freeze-out densities from p-p, \mbox{p-$\alpha$}, and d-$\alpha$ 
correlation functions (open symbols) were analysed
assuming an instantaneous emission of the particles\protect{\cite{Fritz995}}. 
The closed symbols represent
the freeze-out densities extracted from p-p correlation-functions assuming
a Gaussian time distribution. The lines are explained in the text.}
}
\end{minipage}
\end{figure}

\section{DISCUSSION}

The last collision of the two particles within the surrounding nuclear debris determines 
their freeze-out radius and is believed to occur when their mean free path  
gets larger than the system itself: the system becomes below a certain density transparent. 
We used reconstructed residue masses \cite{Gross971} to calculate  the freeze out 
densities (Fig. \ref{densities}). They are lower than the ground-state density 
of nuclei and therefore, show expansion.
The strong variation of the freeze-out density is caused by the variation of the mass
of the spectator. The freeze-out densities shown as open symbols 
were analysed in a previous work \cite{Fritz995} with the assumption that the 
particles are emitted instantaneously. Correcting the source size with help of the 
directional cuts for the emission times of the protons results in densities 
(closed symbols) which are only slightly larger compared to the zero time assumption.
The expected higher freeze-out densities were lowered by two effects, the collective 
movement of the source and by the initial Coulomb correlations within the source. 
In the left panel of Fig.\ref{datasim} we show for \mbox{$20 \le Z_{BOUND} \le 40$} 
the simulation of a purely thermal source (thin lines) with the same radius. 
Guided by the freeze-out conditions of a statistical multifragmentation model 
\cite{Bondorf951} we tried to understand the variation
of the freeze-out density with the impact parameter: assuming a constant crack width
between the fragments, the freeze-out density becomes dependent on the
\mbox{multiplicity $M$}. Hereby, the diameter of a 
nucleus in one dimension is compared to the diameter of an expanded nucleus 
with $M^{1/3}-1$ cracks with the width $2d$. This one dimensional freeze-out radius
is independent on the fragment size and only an approximation in three dimensions. 
In this simplified picture the freeze-out volume over the ground 
state volume is
\begin{math}
V_{freeze}/V_0 = (R_0+d(M^{1/3}-1))^3/R_0^3.
\end{math}
We used the event multiplicity 
from Ref. \cite{Gross971} for different species of fragments. For the calculation of the
primordial multiplicities we assume a simple sequential decay
estimate: we observe in energy spectra for neutrons, protons, and alpha-particles a low
energy component, which for the protons scales with the biggest charge in an event 
\cite{Odeh001} and is likely due to sequential decays. 
We, therefore, select for those particles the high energy component of the spectra.
For the crack width we assume two extreme assumptions\cite{Bondorf8511}: 
for a gas of nucleons the crack width $2d$ is in the order of 1.4 fm, the range 
of nuclear interaction. For an event with heavier fragments the fragments surfaces 
are separated by the distance where the nuclear attraction equals the Coulomb
repulsion \mbox{($2d=2.8$ fm)}. Both assumptions are plotted in Fig.
\ref{densities} as dashed lines. They exhibit the expected decrease
of the freeze-out density with centrality. The extracted proton-proton freeze-out 
densities corrected for emission times follow closely within 20\% 
the predictions for a crack width of $2d=1.4$ fm. 

The other scattering states beside of the most peripheral reactions
are in between of both predictions. These scattering states are, however,
not corrected for emission times. The radius $R_{\tau=0}$ with inclusion
of the emission times $\tau$ can be estimated from the uncorrected radius 
\begin{math}
R_{eff}=\sqrt{R_{\tau=0}^2 + \langle v\tau\rangle^2},
\end{math}
with the thermal velocity $v$ for a fragment with temperature of $T=17$ MeV 
\cite{Odeh001}. This estimate is 
plotted as solid and dashed lines in Fig. \ref{maps} for \mbox{$Z_{BOUND} \le 20$} for
radii from p$\alpha$- and d$\alpha$-correlation functions, respectively. Emission times
between $\tau=10-30$ fm/c are needed to get a freeze-out radius comparable with that 
extracted from the emission-time corrected pp-analysis. 

The times between the emission of protons are in the order of the passing time of the 
projectile through the target. We, therefore, cannot exclude that the protons come
from the fireball, which grows with the centrality of the reaction. Attempts to fit
the energy spectra in Fig. \ref{spectra1} with an fireball localized in momentum space 
failed. 
Nonetheless, the intriguing similarity of the impact-parameter dependence of the 
heavier scattering states (p$\alpha$, d$\alpha$) which differ only by a common factor 
from the pp freeze-out densities indicate the emission from the target spectator.

It is interesting to point out that the average expansion from 
the RMS radius at ground-state to the freeze-out density within our emission times 
is much too high. It is \mbox{$\langle \beta \rangle = 0.19-0.29$ c} for the 
most central bin and disagrees
with the observed energy spectra. A low expansion velocity of about 
\mbox{E/A = 0.5 MeV} (0.03 c), suggested by other studies \cite{Beaulieu007}, 
has the consequence that the nucleus first expands for \mbox{$t\approx 100$ fm/c} 
and then decays within the extracted short times of \mbox{$\tau = 10-20$ fm/c}.   
A reaction where the spectators are diluting each other by scattering out the fireball nucleons
would explain the large radii without substantial amount of expansion. Calculating the 
expansion velocity of a nucleus starting to expand from the ground-state radius of a 
$^{197}Au$ nucleus yields an average expansion velocity of 
\mbox{$\langle \beta \rangle = 0.04-0.09$ c} for the most central bin.
  
\section{CONCLUSION}

We constructed correlation functions for pairs of protons detected at backward angles
in the reaction Au+Au at 1000 AMeV incident beam energy. Using high energy cuts of
\mbox{$E>20$ MeV} we selected protons which are only little affected by sequential 
feeding \cite{Schwarz991}.
Comparing the results with Monte-Carlo simulations within the Koonin-Pratt formalism 
fairly constant freeze-out radii of $R\approx8$ fm are deduced and 
emission times of $\tau=10-15$ fm/c are surprisingly short.
The extracted radii are larger than the ground state
radii of target spectators and show expansion. 
The mass dependence of the decaying target spectator
with impact parameter lead to non constant freeze-out densities which decrease with 
centrality. Employing a multiplicity dependent freeze-out picture 
\cite{Bondorf8511} we are able to describe the absolute value of the freeze-out densities
as well as the impact parameter
dependence. Because of the short emission times in the order
of the passing time of both spectators we cannot exclude that the protons
come from first stage scattering of the nuclear cascade. However the impact 
parameter dependence of the freeze-out densities of heavier scattering states 
from a former analysis \cite{Fritz995,Schwarz991} indicates the emission from the target
spectator. 
A more detailed analysis of these heavier scattering states is necessary to
understand their formation and freeze-out properties.


\end{document}